\documentclass[
superscriptaddress,
 amsmath,amssymb,
 aps,
pra,
]{revtex4-1}

\usepackage{graphicx}% Include figure files
\usepackage{dcolumn}% Align table columns on decimal point
\usepackage{bm}% bold math

\begin{document}

\title{Properties of a Polarization based Phase Operator}

\author{Chandra Prajapati}
\email{chandra9.iitd@gmail.com}
\affiliation{
 Department of Physics, Indian Institute of Technology Delhi, Hauz Khas, 110016, New Delhi, India}%
\author{D. Ranganathan}%
 \email{dilip@physics.iitd.ac.in}
\affiliation{
 Department of Physics, Indian Institute of Technology Delhi, Hauz Khas, 110016, New Delhi, India}%

\date{\today}% It is always \today, today,

%\author{Chandra Prajapati$^{1,*}$ and Dilip Ranganathan$^{**}$}
%\address{$^1$Deptt of Physics, Indian Institute of Technology Delhi\\ Hauz Khas New Delhi 110016, India}
%\address{$^*$Corresponding author: chandra9.iitd@gmail.com \\$^{**}$dilip@physics.iitd.ac.in}

\begin{abstract}
 We define a Hermitian phase operator for zero mass spin one particles (photons) by taking account polarization. The Hilbert space includes the positive helicity states and negative helicity states with opposite circular polarization. We define an operator which corresponds to the physical process of reversing the sense of polarization and acts as a bridge between positive helicity states and negative helicity states.  The exponential phase operator obtained using the entire set is unitary and acts as ladder operator over all the states. The phase operator derived from this exponential operator satisfies the canonical commutation relations with the number operator. We have calculated the density matrix and the phase probability distribution of various states like coherent states, squeezed states and thermal states, to illustrate the utility of our operator. 
\end{abstract}

%\ocis{000.0000, 999.9999.}% REPLACE WITH CORRECT OCIS CODES FOR YOUR ARTICLE
                          % NOTE: \ocis{} IS ALIASED TO \pacs{} BUT MUST
                          % FORMAT THE TERMS CORRECTLY FOR EACH JOURNAL

\maketitle %% null function with osajnl.sty
%\twocolumn
\section{Introduction}
 As almost all measurements on electromagnetic waves are measurements of intensity or phase; there has been an abiding interest in a phase operator for light. Having identified the number operator as the quantity proportional to intensity, Dirac \cite {dirac} decomposed the annihilation operator for the harmonic oscillator as $\hat{a} = \hat{E}\sqrt{\hat{n}}={e^{i\hat{\phi}}}\sqrt{\hat{n}}$. Here $\hat{\phi}$ is the phase operator and $\hat{n}$ the number operator of the oscillator. The number operator was to be the quantum equivalent of the classical action and satisfy canonical commutation relations with the phase, $\left[{\hat{n}}, {\hat{\phi}} \right] = i$. As he pointed out, this phase operator is not well defined as there is a lower bound on the eigenvalues of the number operator. In addition, the phase is periodic and defined modulo $2\pi$.  Subsequent workers \cite{susskind, carruthers,newton, garrison, barnett, pegg, luis, bergou, vourdas, popov} all had the same problems of either lack of unitarity of the exponential operator or failure to obtain the canonical commutation relations for the number operator with the phase operator. An excellent review of the above ideas as well as those to be described below was given by Lynch \cite{lynch}.\\ 
 
  Pegg and Barnett \cite{barnett, pegg} suggested the most popular way round these difficulties. The idea is to truncate the oscillator states to a finite number say $s$. The operators in the resulting finite dimensional Hilbert space are well behaved, this is extremely useful as a practical tool. However, questions have been raised \cite{luis, bergou, vourdas} of whether such a weak convergence to the limiting infinite dimensional operator gives the same results as the calculations directly using infinite dimensional operators. In view of these difficulties, Noh,  Fougeres and Mandel \cite{noh, noh1} even suggested that a unique phase operator should not be sought. They suggest that we should only use an operational definition of phase based on experiment performed.\\
  
  With the recent development of quantum tomographic methods to study entanglement and deduce the Wigner distribution, the phase or more accurately the phase difference has again become very topical \cite{braunstein}. There has been many subsequent efforts to define a phase operator both for photons and for the closely related problem of the Harmonic oscillator \cite{ban, daoud, xian, hradil, garcia, shunlong, bjork}. Many of these, especially the work by Ban, Luis and Hradil, use products of two boson operators which are especially well suited to study parametric processes and higher order interference but are also used to cover the case of two orthogonal light polarization modes. These two boson representations are best suited to treat problems where the total number of photons is constant, as the two boson representations used are isomorphic to the angular momentum group. However as pointed out by Bjork et. al., \cite{bjork} these are not well defined in the vacuum state and indeed require a separate definition of the two mode vacuum state, which is annihilated by all elements of the Schwinger two boson algebra.  Further, the addition of one more photon which can take place spontaneously or by stimulated emission, with very different consequences for the phase of the state is not naturally incorporated in the theory.
  
  There is another representation of the electromagnetic field, with the photon helicity as an internal quantum number, which was advocated by Bialynicki-Birula \cite{bialynicki} and independently by Sipe \cite{sipe}, this suggestion made in the context of the photon wave function was subsequently shown to yield many other properties of the electromagnetic field in a fashion such that the passage from the classical theory to the quantum electrodynamics is particularly transperant, \cite{bialynicki1, raymer, smith, birula, saldanha}. Our representation of the field in this paper is also over $2\otimes(0, \infty)$, rather than the usual product state formalism.
  
  The Hilbert space we consider is a tensor product of a two internal dimensional space of polarization and the Fock space of a single polarization mode. This agrees with the representations found by Wigner, \cite{bargmann, wigner} for massless relativistic particles. Bargmann and Wigner showed the appropriate relativistic representation for photons is two dimensional, this was  elaborated later \cite{kim} to show that it is indeed the natural representation to take into account all polarization phenomena. As the detailed calculations of Kim et. al. show the operators for the commonly used devices like polarizers, phase plates are particularly simple in this representation being constant matrices. The action of such operators is similar on states with the same polarization but different number of photons.
  
  As the Riemann-Silberstein vector $\vec{E} + i \vec{B}$ as the photon wavefunction advocated by Bialynicki-Birula and Sipe yields a representation for photons in both position and momentum space, this method can be also used to study nonlocal correlations of the EPR type, which are not easily defined in the usual product space representations. Also it is known that the the Cartesian components of the light polarization need to commute to give a definition consistent with Maxwell's equations \cite{li, vanenk, vanenk1}.

Sperling and Vogel \cite{sperling} recently pointed out that removal of a right circularly polarized photon is the same as the addition of a left circularly polarized photon. They then suggest treating the left circularly polarized photons as antiparticles of the right circularly polarized photons and used this to construct a unitary Susskind Glogower operator $\hat{E}$. The difficulty with this theory is that there is a difference in angular momentum between the vacuum for left circularly polarized photons and that for right circularly polarized photons.

We suggest a modification of the Sperling-Vogel procedure, based on two requirements.

First, the quantum phase being an angle,\textit{ the corresponding momentum must be an action variable}.  The number operator acting on a Fock state then measures the number of action quanta in the state. No experiment directly measures the action. However, the number of action quanta is equal to the number of azimuthal angular momenta in that state, which is measurable; so we have a suitable proxy. The essence of our suggestion is that action and phase form a conjugate pair while energy and time form another conjugate pair and treating them as freely exchangeable may be part of the reason for the earlier difficulties.

Second, by passage through a phase plate, it is possible to transform a state of definite momentum and polarization ``a mode'', into one of the orthogonal polarization, ``an orthogonal mode'', while preserving the phase or changing it trivially. \textit{ So  the Hilbert space relevant to phase measurements is that spanned by both the polarization modes taken together.}

 If the number of right circular polarized quanta present is denoted by $ n_{+}  \geq 0$ then the number of left circular quanta present is obviously $ n_{-} \leq 0 $ or vice versa.  This method of labelling of the Helicity eigenstates is consistent with the requirement that if the right circularly polarized photons have positive values of azimuthal polarization and positive action, then the left circularly polarized photons have negative values of azimuthal angular momentum and consequently negative values of action. Our method of labelling permits a natural extension to cover the negative integers required to attain unitarity. Newton \cite{newton}, suggested the use of such a two dimensional representation, but in terms of energy eigenstates, consequently he had to suppress the negative energy states which were only a mathematical intermediate step in his model. The angular momentum of light both orbital and spin has been measured and shown to add algebraically \cite{allen}.

In the next two sections, the Hilbert space and the Susskind Glogower operator are defined  and the operator is shown to be unitary,  while the corresponding phase  operator is Hermitian. We then find the eigen states of the Susskind Glogower operator and show they are orthogonal but non-normalizable. All the previous results cited above can be recovered by restricting our space to states of a single polarization. While we outline our results using the two opposite circular polarizations, they hold in any polarization basis.  Finally, using the projectors on to the different polarizations we construct operators which change the number of photons in one polarization state while leaving the other unchanged. As this operator for photon removal (addition) commutes with the corresponding annihilation (creation) operator acting on the whole space, we can use these to construct any arbitrary photon state.

\section{The Phase Operator}
The positive frequency components of a single momentum mode of the electromagnetic field are usually written as \cite{mandel},
\begin{equation}
\label{field}
\hat{E}(\vec{r},t) \propto \hat{a}_{+} e^{i({\vec{k}\cdot\vec{r}-\omega t})}+{\hat{a}_{-}} e^{i({\vec{k}\cdot\vec{r}-\omega t})}.
\end{equation}

Here $\hat{a}_+$ is the annihilation operator for right circularly polarized light, and ${{\hat{a}}}_-$ is the annihilation operator for left
circularly polarized light.  These operators satisfy the commutation relations,
\begin{eqnarray}
\left[\hat{a}_{+},{{\hat{a}}^{\dagger}}_{+}\right] =\left[\hat{a}_{-},{{\hat{a}}^{\dagger}}_{-}\right]=1, \nonumber \\
\left[\hat{a}_{+},{\hat{a}}_{-}\right] =\left[\hat{a}_{+},{{\hat{a}}^{\dagger}}_{-}\right]=0.
\end{eqnarray} 
If $\hat{P}$ is the helicity operator then 
\begin{equation}
\label{eq}
\hat{P}|n_{\pm}>=\pm|n_{\pm}> \;  \mathrm{and} \;   (\hat{P})^2= 1
\end{equation}
As there are two states for a given momentum $k$ and these are orthogonal (Malus law) the natural representation appears to be in terms of a two-dimensional \textit{internal} space. 
\begin{equation}
\left[\begin{array}{c}
%\label{decomp}
 |n_{+}> \\ |n_{-}> 
 \end{array} \right] = \left[\begin{array}{c}
 |n_{+}> \\ 0 
 \end{array} \right] + \left[\begin{array}{c}
 0 \\ |n_{-}> 
 \end{array} \right],
\label{state-space}
\end{equation}
along with
\begin{eqnarray}
\label{counter}
\hat{P}\left[\begin{array}{c} |n_{+}> \\ 0 \end{array}\right] = +\left[\begin{array}{c} |n_{+}> \\ 0 \end{array}\right],\nonumber \\
\hat{P}\left[\begin{array}{c} 0\\ |n_{-}>  \end{array}\right] = -\left[\begin{array}{c} 0\\ |n_{-}>  \end{array}\right].
\end{eqnarray}
 We use the operator ${\hat{P}} $ to define operators which project into the circular polarization eigenstates.
\begin{eqnarray}
\label{projectors}
\left(\frac{I+\hat{P}}{2}\right)\left[\begin{array}{c} |n_{+}> \\ |n_{-}> \end{array}\right] = \left[\begin{array}{c} |n_{+}> \\ 0 \end{array}\right], \nonumber\\
\left(\frac{I-\hat{P}}{2}\right) \left[\begin{array}{c} |n_{+}> \\ |n_{-}> \end{array}\right] = \left[\begin{array}{c} 0 \\ |n_{-}>\end{array}\right].
\end{eqnarray}
To incorporate the polarization directly into our description, we note that the commutation relations for the annihilation and creation operators imply that if $|n>$ is an eigenstate of the number operator with eigenvalue $n$ then
\begin{eqnarray}
\label{qnumbers}
{\hat{a}}^{\dagger} \hat{a} (\hat{a}|n>) &=& (n-1)(\hat{a}|n>),\nonumber\\ 
{\hat{a}}^{\dagger} \hat{a} ({\hat{a}}^{\dagger}|n>) &=& (n+1)({\hat{a}}^{\dagger}|n>). 
\end{eqnarray}
Both are eigenstates with eigenvalues $n-1$ and $n+1$. \textit{The commutation relations impose no further restrictions on $n$}. The series termination requirement yields the fact that $n$ must be an integer. However the set of all non-negative integers as well as the set of all non-positive integers both satisfy equation (\ref{qnumbers}).
\begin{eqnarray}
{\hat{a}}_{+} |n>& = &\sqrt{n}\,|n-1>,\nonumber \\
{{\hat{a}}_{+}}^{\dagger} |n> &= &\sqrt{n+1}\,|n+1>,\nonumber \\
{\hat{a}}_{+} |0> = 0,\;\; &\Rightarrow& n = 0, +1,+2,+3 \dots, \\
{\hat{a}}_{-} |n> &=& \sqrt{|n|}\,|n -1>,\nonumber \\
{{\hat{a}}_{-}}^{\dagger} |n> &=& \sqrt{|n+1|}\,|n+1>, \nonumber \\
{{\hat{a}}^{\dagger}}_{-} |-1> = 0,\;\; &\Rightarrow& n = -1,-2,-3 \dots .
\end{eqnarray}
We now apply this result to our formulation; the modified annihilation operator of system is written as 
\begin{equation}
\label{eq}
\hat{a}_{m}=\left[\begin{array}{cc}
\hat{a}_{m+} & 0\\
\hat{a}_{v} & \hat{a}^{\dagger}_{m-}
\end{array}\right].
\end{equation}
 Where $\hat{a}_{m+}$ acts on positive states and $\hat{a}_{m-}$ acts on negative states and are given by, 
\begin{eqnarray}
\label{modified}
\hat{a}_{m+}=\hat{P}\left(\frac{1+\hat{P}}{2}\right)\hat{a}_{+},\;      
\hat{a}_{m-}=\hat{P}\left(\frac{1-\hat{P}}{2}\right)\hat{a}^{\dagger}_{-},\nonumber \\
\hat{a}_{v}=\left(\frac{1-\hat{P}}{2}\right)|-1><0|\left(\frac{1+\hat{P}}{2}\right),\nonumber \\
\left[\hat{a}_{+},\hat{P}\right]=0  ,\;  \left[\hat{a}_{-},\hat{P}\right]=0. 
\end{eqnarray}
 For positive helicity states $\hat{a}_{+}$ is the annihilation operator and for negative helicity states $\hat{a}^{\dagger}_{-}$ is the annihilation operator and $\hat{a}_{v}$ is an operator which serves as bridge between two subspaces of positive helicity states and negative helicity states. It causes the vacuum of positive helicity states to change into vacuum of negative helicity states and vice-versa. Note that quantum theory requires them both to have the same energy at least in the absence of a medium.
 
 The state space is spanned by the basis,
\begin{equation}
\left[\begin{array}{c}
 |n_{+}>\\ 0
  \end{array}\right]
\end{equation}
and
\begin{equation}
\left[\begin{array}{c}
0 \\|n_{-}-1>
\end{array}\right]
\end{equation}
with,
\begin{eqnarray}
\hat{a}_{m+}|n_{+}> =\sqrt{n_{+}}|n_{+}-1>   ,&\;\;\; &  \hat{a}_{m+}|n_{-}>=0   ,\nonumber \\
  n_{+}&=&0,1,2,3,\dots \nonumber \\
\hat{a}^{\dagger}_{m-}|n_{-}>=\sqrt{\left|n_{-}\right|}|n_{-}-1>  ,&\;\;\;&  \hat{a}^{\dagger}_{m-}|n_{+}>=0  ,\nonumber \\ 
 n_{-}&=&-1,-2,-3,\dots .
\end{eqnarray} 
 Thus our annihilation and creation operators are
\begin{equation}
\hat{a}_{m}=\left[\begin{array}{cc}
\hat{a}_{+} & 0\\
 |-1><0| & \hat{a}_{-}\end{array}\right],
 \end{equation}
 and
 \begin{equation}
\hat{a}^{\dagger}_{m}=\left[\begin{array}{cc}
\hat{a}^{\dagger}_{+} & |0><-1|\\
 0 & \hat{a}^{\dagger}_{-}\end{array}\right].
\end{equation} 
 Using these annihilation and creation operators, number operator and an exponential of the phase operator are constructed in a similar way as done by Susskind and Glogower. 
 The number operator  in this representation is,
\begin{equation}
%\begin{array}{ccc}
\hat{n}_{m}=\left[\begin{array}{cc}
\hat{n}_{+} & 0 \\ 0 & \hat{n}_{-}-1 \end{array}\right],
\end{equation}
and the exponential of the phase operator
\begin{equation}
%\nonumber \;\;\mathrm{and} \nonumber \\
\hat{E}_{m}=\hat{a}_{m}\frac{1}{\sqrt{\hat{n}_{m}}}=\left[\begin{array}{cc}
\hat{a}_{+}\frac{1}{\sqrt{\hat{n}_{+}}} & 0 \\
\hat{a}_{v} & \hat{a}_{-}\frac{1}{\sqrt{\hat{n}_{-}}}\end{array}\right]
%\end{array}
\end{equation}
The exponential of phase operator was found to act as ladder operator over all the states.
\begin{eqnarray}
\hat{E}_{m}|n_{m}>=|n_{m}-1> \:, \; \forall \; n_{m}=n_{+} \;\mathrm{and} \;n_{-}, \nonumber \\
\hat{E}_{m}|0>=|-1> \; \;\; \mathrm{and} \;\;\; \hat{E}^{\dagger}_{m}|-1>=|0>.
\end{eqnarray}
The exponential of phase operator is unitary as,
\begin{equation}
\hat{E}_{m}\hat{E}^{\dagger}_{m}=\hat{E}^{\dagger}_{m}\hat{E}_{m}=\hat{1}.
\end{equation}
so we can write $\hat{E}_{m}= e^{ i \hat{\phi}_{m}}$ and hence the phase operator is Hermitian and satisfies the canonical commutation relations with number operator as
\begin{equation}
\left[\hat{E}_{m}, \hat{n}_{m}\right]=\hat{E}_{m} \;\;  \mathrm{so} \;\; \left[\hat{\phi}_{m}, \hat{n}_{m}\right] =-i.
\end{equation}
Therefore, the canonical representation for the number operator in the basis of phase ($\phi$) eigenstates is written as 
\begin{equation}
\hat{n}_{m}=i \frac{\partial}{\partial{\phi}} .
\end{equation}
 The commutation relation of our modified annihilation and creation operators becomes, 
\begin{equation}
\left[\hat{a}_{m}, \hat{a}^{\dagger}_{m}\right]=\left[\begin{array}{cc}
\hat{P}-|0><0| & 0 \\ 0 & \hat{P}+|-1><-1| \end{array}\right].
\end{equation}
 So the non-diagonal matrix elements vanish and diagonal matrix elements give +1 in the positive number states and -1 in the negative number states as required. Most interestingly they vanish in both vacuum states
\begin{eqnarray}
<n_{+}| 
\left[\hat{a}_{m}, \hat{a}^{\dagger}_{m}\right]
 |n_{+}> = 1 ,
 \nonumber \\
<n_{-}|\left[\hat{a}_{m}, \hat{a}^{\dagger}_{m}\right]|n_{-}> = -1, \nonumber \\
<-1|\left[\hat{a}_{m}, \hat{a}^{\dagger}_{m}\right]|-1> = <0|\left[\hat{a}_{m}, \hat{a}^{\dagger}_{m}\right]|0> = 0. 
\end{eqnarray}
 This means all the commutation relations vanish between the vacuum states, showing that there is no energy difference between the vacuum states of either polarization. The two vacuum states are degenerate. So the appropriate vacuum state for bosons is their symmetric combination as required by quantum theory. The eigenvalue equation for $\hat{E}_{m}$ gives the phase states
\begin{equation}
|\phi_{m}>=\frac{1}{\sqrt{2\pi}}{\displaystyle\sum_{n=0}^{\infty}}\left[\begin{array}{c} e^{ i (n+\frac{1}{2})\phi} |n>\\ 0 \end{array}\right] %\nonumber \\
+ \frac{1}{\sqrt{2\pi}}\displaystyle\sum_{n=-\infty}^{-1}\left[\begin{array}{c} 0 \\ e^{ i (n+\frac{1}{2})\phi}|n> \end{array}\right].
\end{equation}
 Thus we conclude that the two vacuum states differ in phase and are orthogonal
\begin{eqnarray}
 |\phi_{0}> &=&\frac{1}{\sqrt{2\pi}} e^{ i \phi/2} \left[\begin{array}{c} |0> \\ 0 \end{array}\right] ,\nonumber \\ 
 |\phi_{-1}> &=& \frac{1}{\sqrt{2\pi}}e^{ -i \phi/2} \left[\begin{array}{c} 0 \\ |-1> \end{array}\right] \nonumber \\
  <-1|0> = & 0 & = <0|-1>. 
\end{eqnarray}
\section{Phase Probability Density} 
Using this phase state,  we next calculated probability of phase distribution in coherent states, squeezed states and thermal states. If the density matrix for the pure phase state $|\psi>$  is given by  $\hat{\rho}$, then the phase probability distribution $p_{\hat{\rho}}(\phi)$ is given by
\begin{eqnarray}
p_{\hat{\rho}}(\phi)=<\psi|\hat{\rho}_{\phi}|\psi>   \;\;\;\mathrm{where} \;\;\; \hat{\rho}_{\phi}=|\phi_{m}><\phi_{m}| \nonumber \\
 |\phi_{m}>=\frac{1}{\sqrt{2\pi}}\displaystyle\sum_{n=0}^{\infty}\left[\begin{array}{c} e^{ i (n+\frac{1}{2})\phi}|n>\\ 0 \end{array}\right] + 
 \frac{1}{\sqrt{2\pi}}\displaystyle\sum_{n=-\infty}^{-1}\left[\begin{array}{c} 0 \\ e^{ i (n+\frac{1}{2})\phi}|n> \end{array}\right].
\end{eqnarray}
where $\phi$ is the phase and $n$ is any integer between 0 and infinity.
The density matrix for phase state is given by
\begin{eqnarray}
\hat{\rho}_{\phi}=\frac{1}{2{\pi}}\left( \displaystyle\sum_{n=0}^{\infty}\displaystyle\sum_{m=0}^{\infty}\left[\begin{array}{cc} e^{ i(n-m)\phi} |n><m| & 0 \\  0 & 0 \end{array}\right] + \displaystyle\sum_{n=-\infty}^{-1}\displaystyle\sum_{m=-\infty}^{-1}\left[ \begin{array}{cc} 0 & 0 \\ 0 & e^{ i(n-m)\phi}|n><m|  \end{array} \right] \right. \nonumber\\
   \left. + \displaystyle\sum_{n=0}^{\infty}\displaystyle\sum_{m=-\infty}^{-1}\left[\begin{array}{cc} 0  &  e^{ i(n-m)\phi}  |n><m| \\  0 & 0  \end{array}\right] + \displaystyle\sum_{n=-\infty}^{-1}\displaystyle\sum_{m=0}^{\infty}\left[\begin{array}{cc} 0  &  0  \\  e^{ i(n-m)\phi}|n><m|  & 0  \end{array}\right] \right)
\end{eqnarray}
The corresponding probability density is
\begin{eqnarray}
p_{\hat{\rho}}(\phi)=\frac{1}{2\pi}<\psi|\left(\displaystyle\sum_{n=0}^{\infty}\displaystyle\sum_{m=0}^{\infty}\left[\begin{array}{cc} e^{ i(n-m)\phi} |n><m|  &  0  \\  0 & 0 \end{array}\right] +          
 \displaystyle\sum_{n=-\infty}^{-1}\displaystyle\sum_{m=-\infty}^{-1}\left[\begin{array}{cc} 0  &  0  \\  0 & e^{ i(n-m)\phi} |n><m|   \end{array}\right] \right. \nonumber  \\
  + \left. \displaystyle\sum_{n=0}^{\infty}\displaystyle\sum_{m=-\infty}^{-1}\left[\begin{array}{cc} 0  &  e^{ i(n-m)\phi}|n><m|   \\ 0 & 0 \end{array} \right] + \displaystyle\sum_{n=-\infty}^{-1}\displaystyle\sum_{m=0}^{\infty}\left[\begin{array}{cc} 0  &  0  \\   e^{ i(n-m)\phi} |n><m| & \end{array}\right] \right) |\psi>.
\end{eqnarray}
The first two terms give phase probability distributions of the right circular polarization and left circular polarization and the remaining two terms give the polarization interference. 
\subsection{Phase probability distribution for a coherent state}
Taking account the polarization, the coherent state is written as 
\begin{equation}
|\alpha>= e^{\frac{-{|\alpha|}^2}{2}}\left(\displaystyle\sum_{n=0}^{\infty}\frac{\alpha^n}{\sqrt{n!}} \left[\begin{array}{c}  |n> \\ 0
 \end{array}\right] + \displaystyle\sum_{n = 0}^{\infty}\frac{\alpha^n}{\sqrt{n!}} \left[\begin{array}{c}  0 \\ |-n-1>
 \end{array}\right]\right). 
\end{equation}
Positive value of $n$ represents all right circular polarization states and negative value of $n$ represents all left circular polarization states. Substituting this value in the expression of phase probability density gives, 
\begin{eqnarray}
p_{\rho}(\phi)=e^{-{|\alpha|}^2}\left(\displaystyle\sum_{n=0}^{\infty}\frac{{\alpha^{*}}^{n}}{\sqrt{n!}} \left[\begin{array}{cc}  <n| & 0
 \end{array}\right] + \displaystyle\sum_{n=0}^{\infty}\frac{{\alpha^{*}}^{n}}{\sqrt{n!}} \left[\begin{array}{cc}  0 & <-n-1|
 \end{array}\right]\right)  \nonumber \\
 \times\frac{1}{2\pi}\left(\displaystyle\sum_{n=0}^{\infty}\displaystyle\sum_{m=0}^{\infty}\left[\begin{array}{cc} e^{ i(n-m)\phi}|n><m|   &  0  \\  0 & 0 \end{array}\right] + \displaystyle\sum_{n=0}^{\infty}\displaystyle\sum_{m=0}^{\infty}\left[\begin{array}{cc} 0  &  0  \\  0 & e^{ -i(n-m)\phi} |-n-1><-m-1| \end{array}\right] \right. \nonumber\\
    \left. + \displaystyle\sum_{n=0}^{\infty}\displaystyle\sum_{m=0}^{\infty}\left[\begin{array}{cc} 0  &  e^{ i(n+m)\phi} |n><-m-1|  \\  0 & 0 \end{array}\right] + \displaystyle\sum_{n=0}^{\infty}\displaystyle\sum_{m=0}^{\infty}\left[\begin{array}{cc} 0  &  0  \\  e^{ -i(n+m)\phi} |-n-1><m| & 0  \end{array}\right] \right) \nonumber\\
 \times \left(\displaystyle\sum_{n=0}^{\infty}\frac{{\alpha}^n}{\sqrt{n!}} \left[\begin{array}{c}  |n> \\ 0
 \end{array}\right] + \displaystyle\sum_{n=0}^{\infty}\frac{{\alpha}^n}{\sqrt{n!}} \left[\begin{array}{c}  0 \\ |-n-1>
 \end{array}\right]\right). 
\end{eqnarray}  
 Solving for all the terms and taking $\alpha=1$ we get the phase probability density for right circular and left circular polarization as,
\begin{eqnarray}
p^{R}_{\rho}(\phi)=\frac{1}{2\pi}e^{-|\alpha|^2}\displaystyle\sum_{n=0}^{\infty}\displaystyle\sum_{m=0}^{\infty}\frac{{\alpha^{*}}^{n}{\alpha^{m}}}{\sqrt{n!m!}}e^{ i (n-m)\phi} \nonumber \\
p^{L}_{\rho}(\phi)=\frac{1}{2\pi}e^{-|\alpha|^2}\displaystyle\sum_{n=0}^{\infty}\displaystyle\sum_{m=0}^{\infty}\frac{{\alpha^{*}}^{m}\alpha^{n}}{\sqrt{n!m!}}e^{- i (n-m)\phi}
\end{eqnarray}
while the polarization interference term is
\begin{equation}
p^{I}_{\rho}(\phi)=\frac{1}{2\pi}e^{-|\alpha|^2}\displaystyle\sum_{n=0}^{\infty}\displaystyle\sum_{m=0}^{\infty}\frac{{\alpha^{*}}^{n}\alpha^{m}}{\sqrt{n!m!}}e^{ i (n+m)\phi}\; + \frac{1}{2\pi}e^{-|\alpha|^2}\displaystyle\sum_{n=0}^{\infty}\displaystyle\sum_{m=0}^{\infty}\frac{{\alpha^{*}}^{m}\alpha^{n}}{\sqrt{n!m!}}e^{ -i (n+m)\phi}
\end{equation}
We plot phase probability density of all the polarization states with the phase in the limit $-\pi$ to $+\pi$ taking $\alpha=1$, We get the plots as,

\begin{figure}[!h]
\centering
\includegraphics[width=8.3cm]{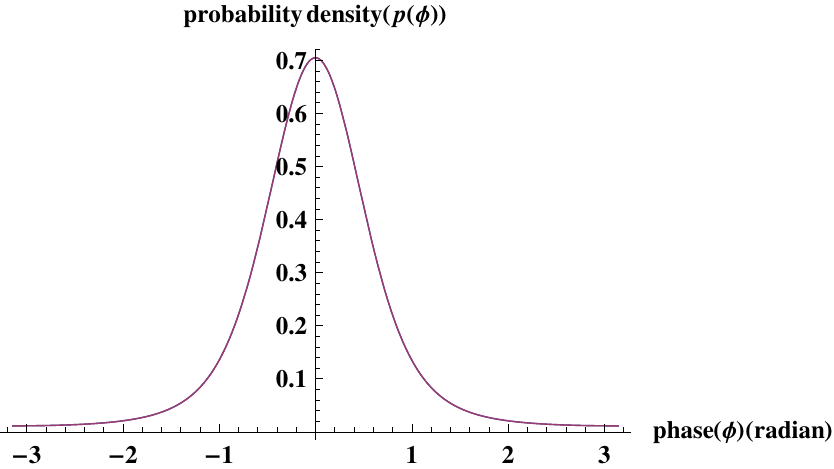} %use this with pdf
\caption{Phase probability distribution of right circular  and left circular polarization for a coherent state which overlap.}
\end{figure}
\begin{figure}[!h]
\centering
\includegraphics[width=8.3cm]{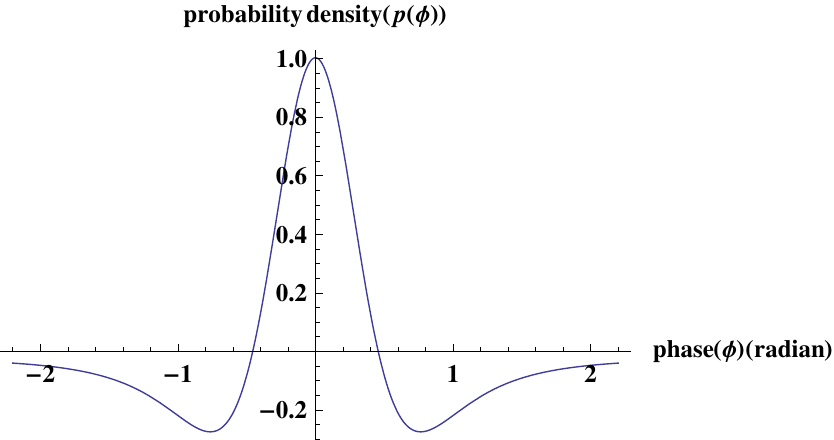} %use this with pdf
\caption{Polarization interference because of right circular and left circular polarization for a coherent state.}
\end{figure}  

\subsection{Phase probability density for a Squeezed state }
The properties of the squeezed states are dependent on the phase so we studied their phase properties. Taking account the polarization, the number states decomposition of squeezed states is written as 
\begin{eqnarray}
|\alpha,\xi>=\frac{1}{\sqrt{\cosh{r}}} e^{\frac{-1}{2} \left[|\alpha|^2+{\alpha^{*}}^{2}e^{ i \theta}\tanh{r}\right]}  
\times \left(\displaystyle\sum_{n=0}^{\infty}\frac{\left(\frac{1}{2}e^{ i \theta}\tanh{r}\right)^{n/2}}{\sqrt{n!}}H_{n}\left[\gamma(e^{ i \theta}\sinh2r)^{-1/2}\right]\left[\begin{array}{c} 1 \\ 0 \end{array}\right]|n> \right. \nonumber\\
\left. + \displaystyle\sum_{n=0}^{\infty}\frac{\left(\frac{1}{2}e^{ i \theta}\tanh{r}\right)^{-n/2}}{\sqrt{n!}}   
\times H_{-n}\left[\gamma(e^{- i \theta}\sinh2r)^{-1/2}\right]\left[\begin{array}{c} 0 \\ 1 \end{array}\right]|-n-1>\right)
\end{eqnarray}
Where $\xi=re^{ i \theta}$ squeeze parameter with $-\infty\leq r \leq\infty$ and $0 \leq \theta\leq 2 \pi$ and $\gamma=\alpha\cosh{r} + \alpha^{*}e^{ i \theta}\sinh{r}$. \\
Putting these values in phase probability density we get for the right circular polarization and the left circular polarization states as
\begin{eqnarray}
p^{R}_{\rho}(\phi)=\frac{1}{2\pi}\frac{1}{\cosh{r}}e^{-\left[|\alpha|^2+{\alpha^{*}}^{2}\cos{\theta}\tanh{r}\right]}
\displaystyle\sum_{n=0}^{\infty}\displaystyle\sum_{m=0}^{\infty}\left(\frac{\tanh{r}}{2}\right)^\frac{m+n}{2}\frac{e^{- i \frac{(n-m)\theta}{2}}}{\sqrt{n!m!}}e^{i (n-m)\phi} \nonumber \\
\times H_{n}\left[\gamma(e^{- i \theta}\sinh2r)^{-1/2}\right]H_{m}\left[\gamma(e^{ i \theta}\sinh2r)^{-1/2}\right] 
\end{eqnarray}
and
\begin{eqnarray}
p^{L}_{\rho}(\phi)=\frac{1}{2\pi}\frac{1}{\cosh{r}}e^{-\left[|\alpha|^2+{\alpha^{*}}^{2}\cos{\theta}\tanh{r}\right]}
\displaystyle\sum_{n=-\infty}^{-1}\displaystyle\sum_{m=-\infty}^{-1}\left(\frac{\tanh{r}}{2}\right)^\frac{m+n}{2}\frac{e^{- i \frac{(n-m)\theta}{2}}}{\sqrt{|n!| |m!|}}e^{i (n-m)\phi} \nonumber \\
\times H_{n}\left[\gamma(e^{- i \theta}\sinh2r)^{-1/2}\right]H_{m}\left[\gamma(e^{ i \theta}\sinh2r)^{-1/2}\right] 
\end{eqnarray}
and for polarization interference
\begin{eqnarray}
p^{I}_{\rho}(\phi)=\frac{1}{2\pi}\frac{1}{\cosh{r}}e^{-\left[|\alpha|^2+{\alpha^{*}}^{2}\cos{\theta}\tanh{r}\right]}
\displaystyle\sum_{n=0}^{\infty}\displaystyle\sum_{m=-\infty}^{-1}\left(\frac{\tanh{r}}{2}\right)^\frac{m+n}{2}\frac{e^{- i \frac{(n-m)\theta}{2}}}{\sqrt{n!|m!|}}e^{i (n-m)\phi} \nonumber \\
\times H_{n}\left[\gamma(e^{- i \theta}\sinh2r)^{-1/2}\right]H_{m}\left[\gamma(e^{ i \theta}\sinh2r)^{-1/2}\right] +  \nonumber \\ \frac{1}{2\pi}\frac{1}{\cosh{r}}e^{-\left[|\alpha|^2+{\alpha^{*}}^{2}\cos{\theta}\tanh{r}\right]}
\displaystyle\sum_{n=-\infty}^{-1}\displaystyle\sum_{m=0}^{\infty}\left(\frac{\tanh{r}}{2}\right)^\frac{m+n}{2}\frac{e^{- i \frac{(n-m)\theta}{2}}}{\sqrt{|n!|m!}}e^{i (n-m)\phi} \nonumber \\
\times H_{n}\left[\gamma(e^{- i \theta}\sinh2r)^{-1/2}\right]H_{m}\left[\gamma(e^{ i \theta}\sinh2r)^{-1/2}\right] 
\end{eqnarray}
We plot phase probability density for all the terms with phase varying from $-\pi$ to $+\pi$, taking the parameters as  $r=1$ and $\theta=0$ and $\alpha=1$, we get the expressions as
\begin{eqnarray}
p^{R}_{\rho}(\phi)=\frac{1}{3.086\pi}e^{(-1.7616)}\displaystyle\sum_{n=0}^{\infty}\displaystyle\sum_{m=0}^{\infty}(0.3808)^{\frac{m+n}{2}}\frac{e^{ i (n-m)\phi}}{\sqrt{n!m!}}H_{n}\left[1.472\right]H_{m}\left[1.472\right], \nonumber \\
p^{L}_{\rho}(\phi)=\frac{1}{3.086\pi}e^{(-1.7616)}\displaystyle\sum_{n=-\infty}^{-1}\displaystyle\sum_{m=-\infty}^{-1}(0.3808)^{\frac{m+n}{2}}\frac{e^{ i (n-m)\phi}}{\sqrt{|n!| |m!|}}H_{n}\left[1.472\right]H_{m}\left[1.472\right], \nonumber \\
p^{I}_{\rho}(\phi)=\frac{1}{3.086\pi}e^{(-1.7616)}\displaystyle\sum_{n=0}^{\infty}\displaystyle\sum_{m=-\infty}^{-1}(0.3808)^{\frac{m+n}{2}}\frac{e^{ i (n-m)\phi}}{\sqrt{n!|m!|}}H_{n}\left[1.472\right]H_{m}\left[1.472\right], \nonumber \\
+\frac{1}{3.086\pi}e^{(-1.7616)}\displaystyle\sum_{n=-\infty}^{-1}\displaystyle\sum_{m=0}^{\infty}(0.3808)^{\frac{m+n}{2}}\frac{e^{ i (n-m)\phi}}{\sqrt{|n!|m!}}H_{n}\left[1.472\right]H_{m}\left[1.472\right].
\end{eqnarray}
\newpage
The plots are, 
\begin{figure}[!h]
\centering
\includegraphics[width=8.3cm]{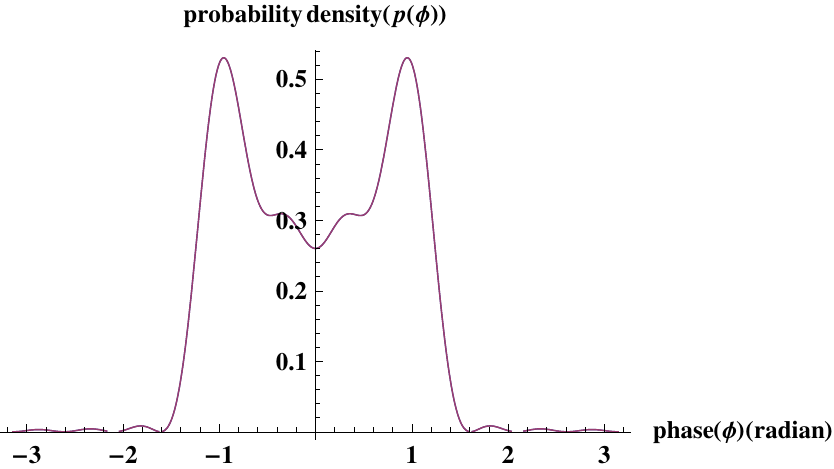} %use this with pdf
\caption{Phase probability distribution of right circular and left circular polarization for a squeezed state which overlap.}
\end{figure}  
\begin{figure}[!h]
\centering
\includegraphics[width=8.3cm]{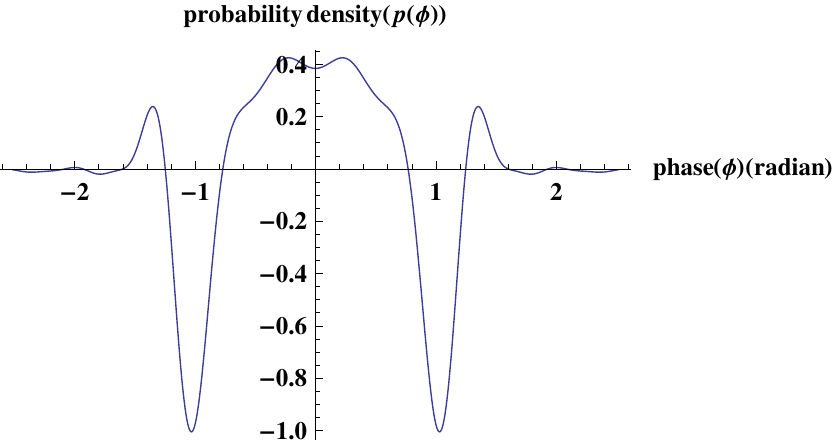} %use this with pdf
\caption{Polarization interference because of right circular and left circular polarization for a squeezed state.}
\end{figure}  
\subsection{Phase probability density for a thermal states}
Thermal states are defined by average photon number and have a completely random phase distributed over the period. The thermal states are a mixture of all the states, therefore, a thermal state is written as
\begin{equation}
|\psi>=\sum_{n}P_{n}|n> = \displaystyle\sum_{n=0}^{\infty}\frac{(\bar{n})^{n}}{(1+\bar{n})^{n+1}}|n> + \displaystyle\sum_{n =0}^{\infty}\frac{(\bar{n})^{n}}{(1+\bar{n})^{n+1}}|-n-1>,
\end{equation}
where $P_{n}=\frac{(\bar{n})^{n}}{(1+\bar{n})^{n+1}},$ is Bose-Einstein weight summed over all the states.\\
Putting these values in phase probability density, we get for circular polarization and opposite circular polarization as,
\begin{eqnarray}
p^{R}_{\rho}(\phi)=\frac{1}{2\pi(1+\bar{n})^2}\displaystyle\sum_{n=0}^{\infty}\left(\frac{\bar{n}}{1+\bar{n}}\right)^{2n}, \nonumber \\
p^{L}_{\rho}(\phi)=\frac{1}{2\pi(1+\bar{n})^2}\displaystyle\sum_{n=0}^{\infty}\left(\frac{\bar{n}}{1+\bar{n}}\right)^{2n}
\end{eqnarray}
and for polarization interference
\begin{equation}
p^{I}_{\rho}(\phi)=\frac{1}{2\pi(1+\bar{n})^2}\displaystyle\sum_{n=0}^{\infty}\left(\frac{\bar{n}}{1+\bar{n}}\right)^{2n}e^{2 i n\phi} + \frac{1}{2\pi(1+\bar{n})^2}\displaystyle\sum_{n=0}^{\infty}\left(\frac{\bar{n}}{1+\bar{n}}\right)^{2n}e^{-2 i n\phi}
\end{equation}
We plot the phase probability density of all the terms with phase varying from $-\pi$ to $+\pi$, taking the parameter $\bar{n}=1$ 
\begin{figure}[!h]
\centering
\includegraphics[width=8.3cm]{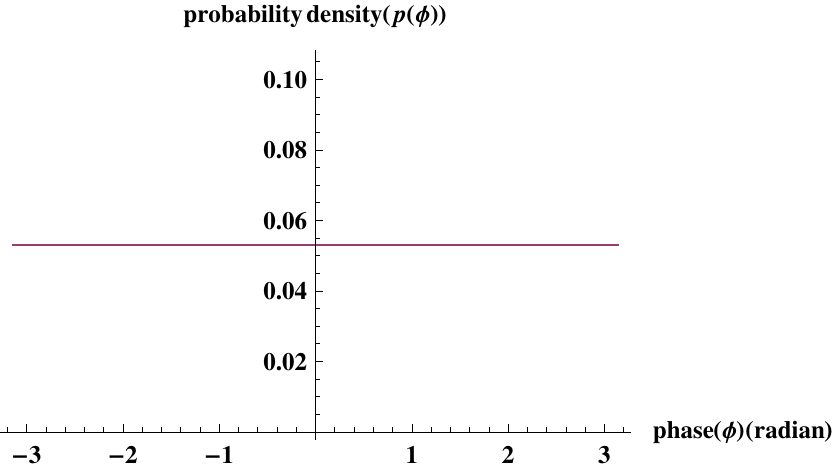} %use this with pdf
\caption{Phase probability distribution of right circular and left circular polarization for thermal state, which overlap.}
\end{figure}  
\begin{figure}[!h]
\centering
\includegraphics[width=8.3cm]{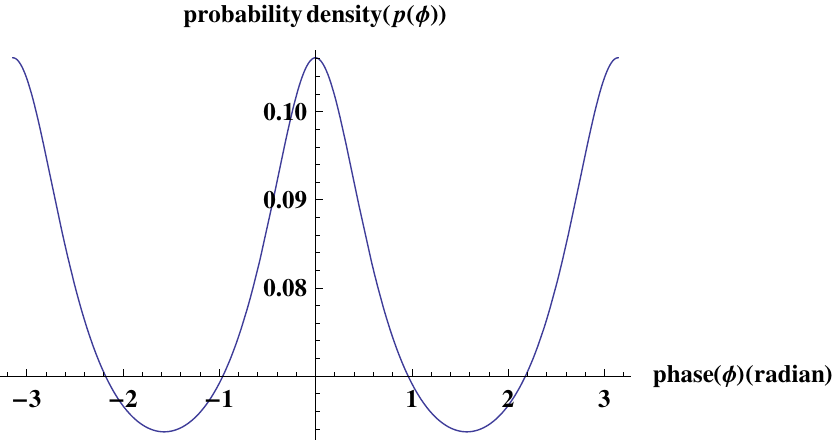} %use this with pdf
\caption{Polarization interference because of right circular and left circular polarizations for the thermal state.}
\end{figure}

\subsection{Conclusion}

We have used the two component form of the Maxwell equations first derived by Wigner \cite{wigner} and subsequently studied in detail by Bialynicki-Birula, Sipe and others to construct a phase operator for photons, which is the canonical conjugate of the number operator. To do so we introduced an operator corresponding to the physical process of reversing the sense of circular polarization. Our method is sufficiently general to be extended to the case of linear isotropic and anisotropic media. As the states used here correspond to the two polarization states studied by Wigner and Kim \cite{kim}, consequently, all the operators derived by them to represent phase plates, attenuators, polarizers etc., can be directly applied to our case without any changes whatsoever even in the limit of vanishing photon numbers. It is interesting to note that it is in this limit of the double vacuum state that the Heisenberg algebra is replaced by complete commutation. Garrison \cite{garrison} had suggested that these operators should not satisfy the commutation relations in the strong (Heisenberg) sense but only in the weaker (Weyl) limit as otherwise the number phase algebra will be the same as the position momentum algebra and this would contradict the requirement that phase has a $2\pi$ periodicity. These methods are best suited for the study of phase and polarization effects in anisotropic and conducting media as these absorptive in nature and the elegance of the conventional Schwinger representation has to be augmented by phenomenological decay terms, while in the wave function representations advocated here, single boson annihilation operators are just as natural as the two boson operators. The other great advantage of the Bialynicki-Birula, Sipe representation is that they permit a position representation for photons, which is consistent with the Newton-Wigner theorem \cite{newton-wigner}. This permits the investigation of phenomena like EPR and other correlations \cite{smith} in a intuitive and physically understandable model. As such experiments are always measurements of polarization coupled  with position our methods should find applications there too.
\subsection{Acknowledgement} Chandra Prajapati  acknowledges the UGC, Government of India for financial support.

%\section*{List of Figure Captions}

%\noindent Fig. 2. ...
%\noindent Fig. 3. ...

%\listoffigures

%\clearpage

%% sample sizing command; other sizing commands (and graphics packages) may be used as well

%  \begin{figure}[htbp]
%  \centering
%  \includegraphics[width=8.3cm]{OT10000F1.eps}
%  \caption{Multipanel figure assembled into one EPS file with proper arrangement and labeling. AO10000F1.eps.}
%  \end{figure}
\section{\label{sec:references}References}

\end{document}